\documentclass[aps,prl,floatfix,showpacs,twocolumn,amsmath,amssymb,superscriptaddress]{revtex4-2}

\usepackage{upgreek}
\usepackage{hyperref}
\usepackage{siunitx}
\usepackage{amsmath}
\usepackage{graphicx}
\usepackage{float}

\begin{document}

\title{Single-beam grating-chip 3D and 1D optical lattices}

\author{Alan Bregazzi}
\author{James P. McGilligan}
\author{Paul F. Griffin}
\author{Erling Riis}
\author{Aidan S. Arnold}
\affiliation{SUPA and Department of Physics, University of Strathclyde, G4 0NG, Glasgow, United Kingdom}

\begin{abstract}

Ultracold atoms are crucial for unlocking truly precise and accurate quantum metrology, and provide an essential platform for quantum computing, communication and memories. One of the largest ongoing challenges is the miniaturization of these quantum devices. Here, we show that the typically macroscopic optical lattice architecture at the heart of many ultra-precise quantum technologies can be realized with a single input laser beam on the same diffractive chip already used to create the ultracold atoms. Moreover, this inherently ultra-stable platform enables access to a plethora of new lattice dimensionalities and geometries, ideally suited for the design of high-accuracy, portable quantum devices.

\end{abstract}

\maketitle

Optical lattices, formed by the standing wave of two or more interfering laser beams are an essential platform for many quantum optics experiments \cite{Dipole_review_Weidemuller,Bloch2008, Optical_clock_review_Ludlow2015}. 
The extremely long atomic coherence times possible in such optical traps have been used extensively in some of the most sensitive measurements to date, notably extending the lifetime in atomic memories many orders of magnitude to several seconds \cite{Dudin2013}, and allowing optical lattice clocks \cite{Takamoto2005} to reach fractional frequency accuracies at the $10^{-18}$ level \cite{Transportable_lattice_clock_Takamoto2020,Sr_lattice_Gravitational_redshift_Bothwell2022}. Such clocks are typically built on 1D lattices, but 3D \cite{Akatsuka2008,Campbell2017} and 2D lattices \cite{Swallows2011} can also be used. 

Lattices in one \cite{Anderson1998}, two \cite{Paredes2004,Kinoshita2004},  and three \cite{Greiner2002} dimensions were also vital components for many key experiments with quantum degenerate gases. By selectively preparing a single 2D layer from such 3D lattices, quantum gas microscopes can be prepared in bosonic \cite{Bakr2009,Stefan_lattice_microscope_Sherson2010} and fermionic \cite{Cheuk2015,Haller2015,Parsons2015} atoms, in an increasing range of atomic species \cite{Yamamoto2016,Tarruell2023,Su2023,Sohmen2023}. Optical lattices for storage, augmented by optical tweezers for transport \cite{Young2022}, can also be used in the burgeoning field of neutral-atom quantum computing \cite{Ebadi2022,Graham2022}.

In recent years there has been a concerted effort to develop portable cold-atom sensors, translating the excellent performance of lab-based experiments to real-world applications. 
Optical-lattice-based devices 
utilise laser-cooled atomic clouds, created in magneto-optical traps (MOTs \cite{Raab1987}) using overlapping near-resonant laser beams,  
which are subsequently loaded into the lattice potentials. The lattices themselves are typically generated from a single laser at a far-detuned 
wavelength which is then split into multiple intersecting laser paths \footnote{Often with frequency shifts between non-colinear beams.}, 
adding more complexity to beam delivery. Mirrors aid stability in 1D lattices, by providing a fixed node of the optical standing wave, however this stable point is often many centimetres away from the atoms. Multiple intersecting 1D lattices can generate higher-dimensional lattices, if they are carefully mutually orthogonally polarised and/or frequency shifted to avoid optical interference which can shake the lattice or even change its geometry \cite{Hopkins1997}.

Previous work on atom chips has demonstrated 
atomic loading from a MOT into a 1D optical lattice by way of a magnetic surface trap \cite{Atom_chip_lattice_Gallego2009,Atom_chip_lattice_Straatsma2015}, however the mirror and 3D MOTs used require four and six beams, respectively.    
The integration of a single-beam pyramid MOT \cite{Lee1996} and 1D and 2D optical lattices  \cite{pyramid_MOT_lattice_NPL_Bowden2019} has also been demonstrated, but the lattice did not use the same input beam direction as the MOT, and  lattice mirrors were $\gg 1\,$mm from the atoms.  
The equivalent to a single-input-beam MOT with a 1D lattice along the same input axis has been achieved in 
interferometry  
\cite{Bodart2010}, however the lattice in this case is used as a beamsplitter instead of storage, and  atom number may be limited \cite{Vangeleyn2009}. 

Grating magneto-optical traps (GMOTs) offer a convenient platform for the formation of ultracold   
atomic clouds with a single input trapping beam \cite{GMOT1_Nshii2013,GMOT2_McGilligan2017}. This technology has now been extended to a range of atomic species \cite{Li_GMOT_Barker2022,Sr_GMOT_NIST_Sitaram2020,Sr_GMOT_PTB_Bondza2022,Sr_grating_design_Burrow2023} and applications, and is well-placed for adoption in compact, cold-atom quantum sensors \cite{GMOT_chip_scale_McGilligan2020,GMOT_interferometer_Lee2022,Abend2023}. 
Here we extend this single-input beam architecture to the formation of optical lattices \cite{GMOT1_Nshii2013}. By aligning a single high intensity far-detuned laser beam onto different areas of the grating chip, 1D, 2D and 3D lattices can be formed -- on the same grating used for the GMOT. All lattice beams share a common nodal plane, pinned to the grating chip surface $\lesssim1\,$mm from the atoms, so the resulting optical lattice is stable against the drift often seen in macroscopic systems. 
We anticipate that this simple, highly compact and stable
optical lattice geometry will be ideal for many portable quantum-optic experiments.

\begin{figure*}[ht]
\centering
\begin{minipage}{.33\linewidth}
\includegraphics[width=\linewidth]{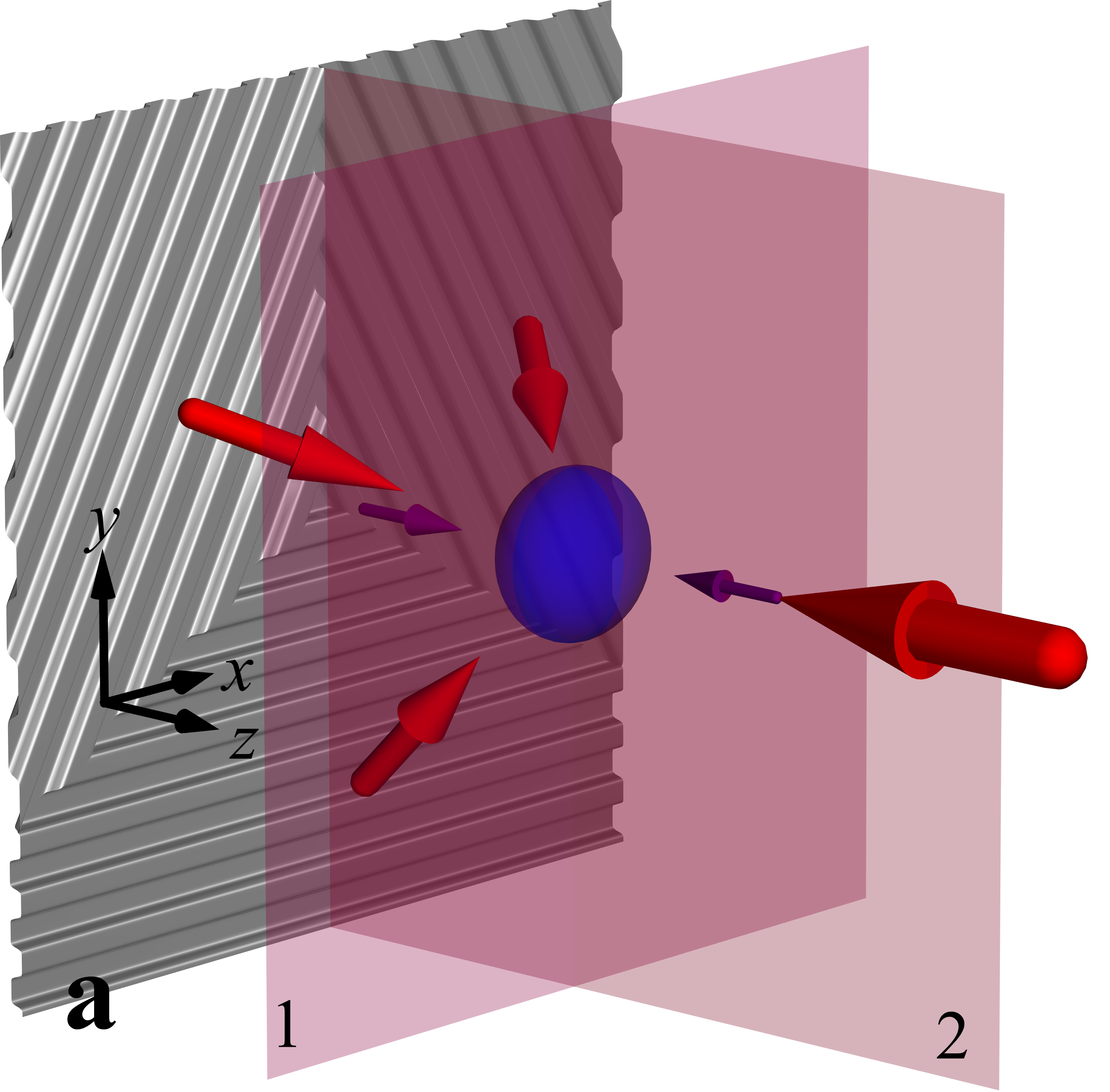}
\end{minipage}
\begin{minipage}{.338\linewidth}
\includegraphics[width=\linewidth]{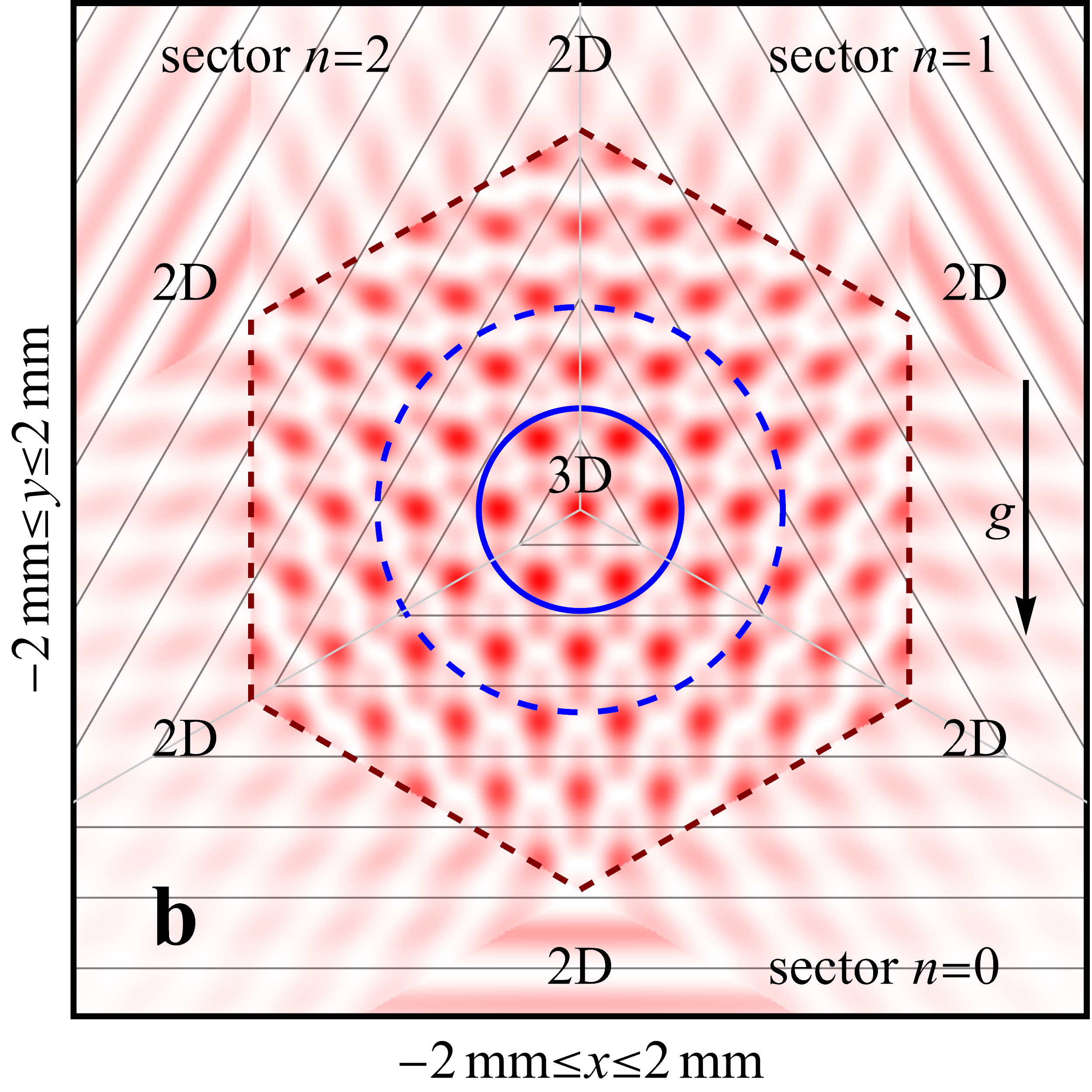}
\end{minipage}
\begin{minipage}{.318\linewidth}
\includegraphics[width=\linewidth]{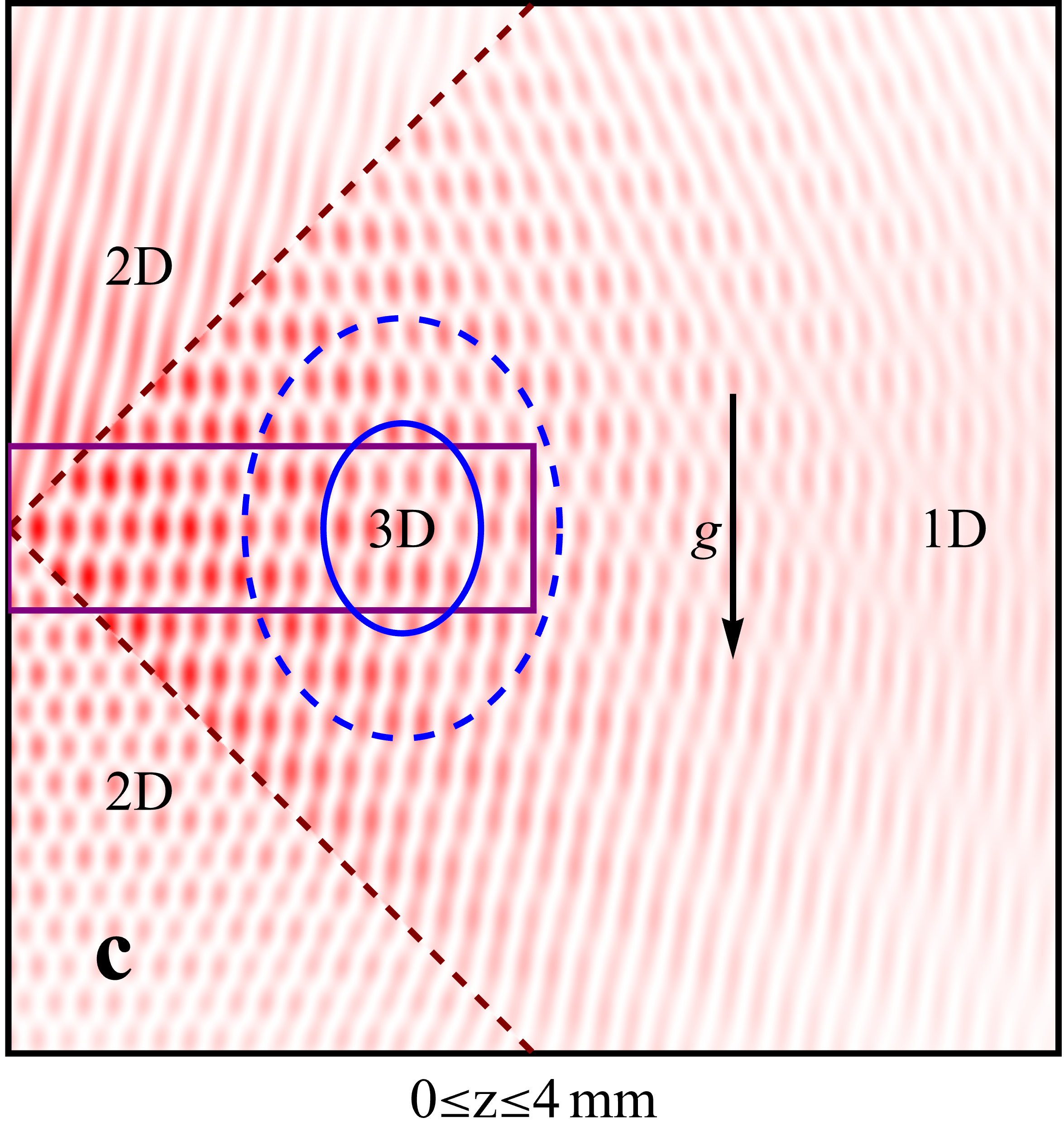}
\end{minipage}
\caption{Grating and lattice geometry in a $(4\,\textrm{mm})^3$ volume, here with both periods magnified by 250 to aid visibility. Red and purple arrows in \textbf{a} indicate the laser beams incident ($-z$ direction) on the triangular grating to generate the 3D and 1D lattice, respectively.  All other arrows ($+z$ direction) indicate principle contributing diffracted orders, and planes 1 and 2 are used to generate images of the 3D lattice structure (red linear shading in \textbf{b} and \textbf{c}, respectively). The gray lines in \textbf{b} are spaced by the grating period.  The 3D lattice is shown in the $xy$-plane at $z=1.5\,$mm above the grating surface (plane 1, \textbf{b}) and  the $zy$-plane at $x=0\,$mm (plane 2, \textbf{c}). The $zx$-plane at $y=0\,$mm is similar to \textbf{c}, and shown in Fig.~SM1 in the Supplementary Material (SM). Dark red dashed lines indicate the 3D beam overlap region boundaries within which the MOT can form, and the solid and dashed blue ellipses indicate the approximately Gaussian MOT's $e^{-1/2}$ and $e^{-2}$ radii, respectively. Gravity $g$ is in the $-y$ direction. Equivalent images of \textbf{c} for the 1D lattice are shown in Fig.~SM2 in the SM.}
\label{fig:fig1_place_holder}
\end{figure*}

Fig.~\ref{fig:fig1_place_holder} \textbf{a} illustrates the central $4\times4\,\textrm{mm}^2$ of the $20\times20\,\textrm{mm}^2$ GMOT chip we use to create a 3D lattice. The chip comprises three sectors of one-dimensional diffraction grating (gray triangles). 
The binary grating sectors have properties and fabrication methods as detailed in \cite{GMOT1_Nshii2013,Cotter2016,McGilligan2016,Sr_grating_design_Burrow2023}, and are optimised for the atomic species utilised here -- $^{87}$Rb.  
We use a similar theoretical description to our MOT modelling papers \cite{Vangeleyn2009,McGilligan2015}, however here we consider the coherent superposition of electric fields rather than the Doppler- and Zeeman-effect sensitive sums of intensity-based average radiation pressures (see Supplementary Material (SM)). Theoretical cross-sections of the 3D lattice itself are shown in Fig.~\ref{fig:fig1_place_holder} \textbf{b} and \textbf{c}, illustrating the different lattice regions with their distinct geometries and dimensions. The 1D lattice model and experimental data are shown in Fig.~SM2 \textbf{a} and \textbf{b}, respectively.

To make the 3D (1D) optical lattices in this paper, we use a single circularly- (linearly-) polarised  laser beam, normally incident near the centre of the grating with wavelength $\approx 780\,$nm ($1070\,$nm),  power $0.9\,$W ($10\,$W), and $e^{-2}$ intensity radius 
$2.5\,$mm ($80\,\upmu$m). For the near-resonant lattice at $\approx 780\,$nm, the laser red-detuning was $78\,$GHz relative to the center of the four Doppler-broadened D$_2$ transitions. Our Al-coated grating, has a material reflectivity of 88\% at $780\,$nm and 95\% at $1070\,$nm \cite{Handbook_of_Optical_materials}. Due to the grating period $d=1080\,$nm, the $780\,$nm light diffracts ($\pm 1$-order) at an angle of $\theta=45^\circ$ to the grating normal with diffraction efficiencies into the first and zeroth orders of $\eta_{\pm1}\approx\,35\,\%$ and  $\eta_{0}\approx\,5\,\%$ respectively \cite{McGilligan2016}. The $1070\,$nm light has most power in the zeroth order $\eta_{0}\approx\,70\,\%$, and the first-order diffraction angle is  large (82$^\circ$) so these beams have no impact on the 1D lattice.

As a basis for loading the optical lattices, $\approx$8$\times10^{6}$ \textsuperscript{87}Rb atoms are loaded into a GMOT and cooled to $\approx$3$\,\mu$K using a standard red molasses cooling stage \cite{GMOT1_Nshii2013,GMOT2_McGilligan2017}. Stray DC magnetic fields are cancelled by three pairs of orthogonal Helmholtz coils. The MOT location with respect to the chip, and its overlap with the optical lattice, can be controlled within the GMOT's optical overlap volume by varying the zero position of the quadrupole magnetic field. 
The typical MOT location and size are also shown in Fig.~\ref{fig:fig1_place_holder} \textbf{b} and \textbf{c}.

The grating chip is bonded to the open end facet of a rectangular cuvette vacuum cell using UHV compatible epoxy, allowing a base vacuum pressure at the $10^{-8}\,$mbar level, as measured by the ion pump. No bowing of the chip is observed due to pressure differential 
\cite{Atom_chip_vacuumSquires2016}. The decision to mount the chip in-vacuum was made to allow the lattice to form as close to the chip surface as possible with no glass interface, enabling future use of atoms in chip-surface traps \footnote{The lattices here could in principle be achieved with an ex-vacuo grating chip, near a thin-walled glass chamber, but this could reduce the lattice capture efficiency.}.

Atom detection is achieved via fluorescence imaging onto a CCD camera sensor. The imaging system is aligned orthogonal to the directions of both the MOT/lattice beam and gravity. 
A spatially selective focal plane is utilised to reduce background noise originating from scattered light \footnote{Additionally, a notch filter with a peak transmission centred around $780\,$nm is used when investigating the 1D $1070\,$nm lattice}. For both the 3D and 1D lattice geometries $150\,\upmu$s long resonant fluorescent images are taken with the MOT light after a $1\,$ms time of flight following the extinction of the lattice beam. 
In both the 3D and 1D lattices, the lattice beam power can be rapidly adjusted by a single-pass acousto-optic modulator (AOM) designed for the beam wavelength. 

The $780\,$nm 3D lattice beam is from a near-resonant diode laser \cite{Thompson2012} with a tapered amplifier. The broadband amplified spontaneous emission (ASE) pedestal associated with this laser produces some near-resonant light which can heat atoms from the lattice via scattering. To reduce this effect we filter the lattice beam through a $7.5\,$cm long $>70\,^\circ$C Rb vapour cell 
\cite{ASE_hot_cell_Daffurn2021}. 
The lattice beam is combined with the MOT beam on a polarising beam splitter.
Atoms are loaded into the 3D lattice potentials by switching the lattice beam on at the end of the red molasses sequence, to mitigate the effect of light shifts on the cooling process. 

To form the far-detuned 1D lattice a $1070\,$nm ytterbium 
laser is used. The vertically polarised light is combined with the MOT light on a dichroic mirror and aligned onto the grating chip. 
Atoms are loaded into the 1D lattice by switching the lattice beam on at the beginning of the red molasses cooling stage.  

\begin{figure}[!t]
\centering
\includegraphics[width=.8\linewidth]{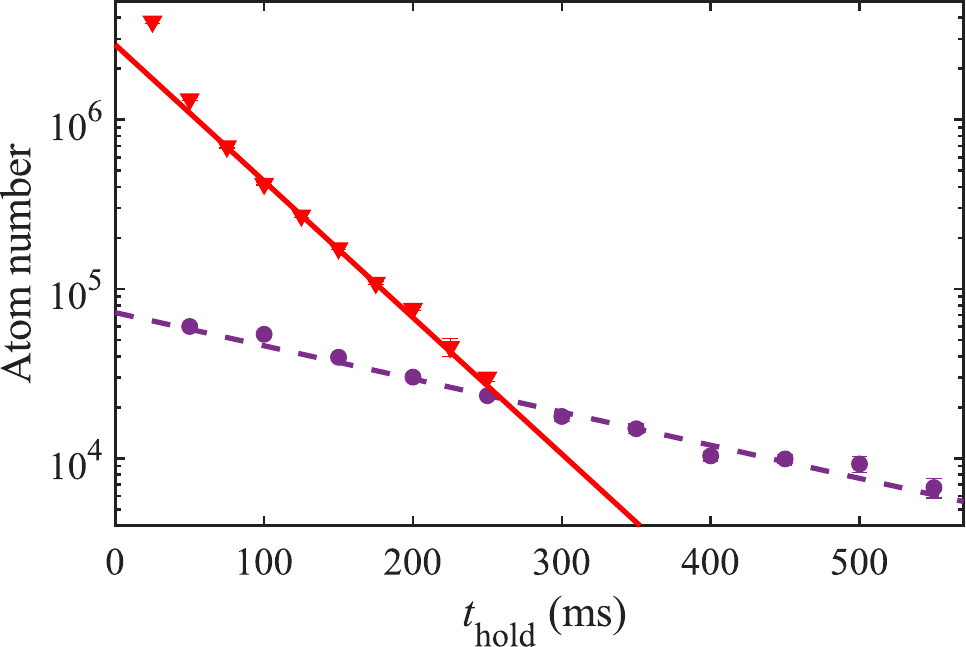}
\caption{ Number of trapped atoms as a function of lattice hold time for the 3D lattice (red triangles, solid fit) and 1D lattice (purple circles, dashed fit). Fitted exponential decays reveal estimated lattice lifetimes of $54\,$ms and $220\,$ms for the 3D and 1D lattice,  respectively.} 
\label{fig:lifetime}
\end{figure}

A log plot of the number of trapped atoms as a function of hold time in both the 3D and 1D lattice is shown in Fig.~\ref{fig:lifetime}. 
Exponential decays are fitted, excluding data where $t_\textrm{hold}<75\,$ms, to  
eliminate untrapped atoms ejected from the imaging region over this time scale due to gravity. The 1D lattice has a much longer lifetime, this is because it is much deeper and the far-off resonance light also induces less scattering and thereby heating than the near-resonant 3D lattice. The overlap volume between the 3D lattice and the MOT is much larger than for the 1D lattice, which can be seen by the difference in initial lattice atom number.

From the time constant of the fits we estimate a 3D lattice lifetime of $54\,$ms.  Due to the limited power available from the laser a relatively low detuning from the Rb D\textsubscript{2} line of $\Delta=-78\,$GHz was used. As a consequence, atomic heating due to photon scattering is significant, estimated to be $1\,$kHz for each photon absorption-emission cycle 
within the deepest (and brightest) parts of the lattice, where the expected potential depth at lattice sites is $40\,\upmu$K. This leads to a maximum heating rate at lattice sites of $10\,\upmu$K/s in good agreement with the observed ensemble heating rate across the lattice of $6\,\upmu$K/s \footnote{Alan Bregazzi, PhD thesis (Strathclyde, 2024).}. Along with any residual unfiltered laser ASE, and the relatively shallow lattice depth, this heating accounts for the shorter near-resonant lattice lifetime.

The far-off resonance 1D lattice is expected to have less photon scattering ($3\,$Hz) with a greater depth at the lattice sites ($500\,\upmu$K). Fitting an exponential decay to lattice atom number data indicates a longer trap lifetime of $220\,$ms. 
An estimation of the background vacuum pressure can be made using the known relation between MOT lifetime and vacuum pressure 
\cite{Arpornthip2012,Moore2015,Burrow2021}. The shallower potential of the 1D lattice compared to the MOT  leads to a shorter background-limited lifetime \cite{GMOT2_McGilligan2017,parametric_oscillation_Wu2006}. When the Rb partial pressure was changed, the MOT fill time was consistently twice as long as the 1D lattice decay time, indicating that both the MOT and 1D lattice are background-pressure limited.  

The long hold time of the atoms, far in excess of that possible if they were in free-fall, is strong evidence of the formation of an optical dipole trap. The observed behaviour could not arise due to  optical molasses or any light scattering forces. Even if we overestimate the effects, by considering a single laser beam scattering at $1\,$kHz against gravity, the maximum acceleration is $6\,$m/s$^2$. Similarly, using a 1D Doppler model with co-propagating beams in the gravity direction with our laser intensity and detuning, the overestimated molasses acceleration reached is a millionth of that due to gravity. 

Although direct imaging of the lattice is not possible without a suitable microscope for site-resolved imaging, 
additional evidence can be obtained via investigating parametric resonances at the lattice trapping frequencies  \cite{CO2_1D_lattice_Friebel1998,parametric_oscillation_Wu2006}. This resonance technique relies on modulating the lattice potential. When the modulation frequency is equal to twice the trapping frequency $\nu_\textrm{trap}$, the kinetic energy of the atoms increases exponentially, resulting in parametric excitation to higher vibrational states of the trapping potential. Eventually atoms are ejected entirely from the trapping potential. This process can be easily observed experimentally by an increase in cloud size, due to atoms entering higher vibrational states, or a drop in atom number as atoms are ejected from the trap. In addition to the fundamental parametric resonance at $2\nu_\textrm{trap}$, sub-harmonics can also be excited at frequencies given by $2\nu_\textrm{trap}/n$, where $n$ is an integer \cite{CO2_1D_lattice_Friebel1998}.

We modulated the lattice beam power in the range $(88-100)\,\%$ using the AOM. In the 3D lattice, modulation is applied for $50\,$ms after first holding the atoms in the static lattice for $50\,$ms to allow for free evaporation \cite{Dipole_review_Weidemuller}. The atoms are then imaged after a $1\,$ms time of flight. In the 1D lattice case, modulation is applied for $150\,$ms due to the improved lattice lifetime and the need to eject a significant number of atoms from the deeper 1D potentials in order to observe the trap resonances. In all cases the modulation scans are randomised in frequency to negate any effects due to long-term drifts in experimental conditions during the data campaigns.

\begin{figure*}[!t]
\centering
\includegraphics[width=\linewidth]{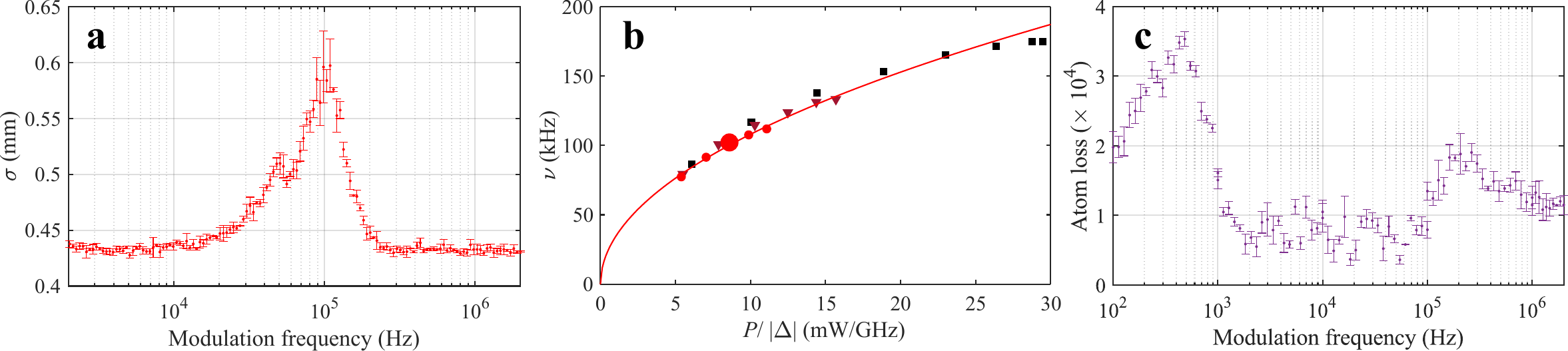}
\caption{(\textbf{a}) Experimental  standard deviation $\sigma$ of the 3D lattice cloud distribution, showing the spectrum of vibrational heating ($\sigma>0.44\,\textrm{mm}$), with beam power $650\,$mW and detuning $\Delta=-78\,$GHz. (\textbf{b}) Primary vibrational resonance frequency plotted as a function of optical power divided by detuning from the \textsuperscript{87}Rb D\textsubscript{2} transition. Black squares $\Delta=-30\,$GHz, dark red triangles $\Delta=-54\,$GHz, light red circles $\Delta=-78\,$GHz. Large light red circle corresponds to the primary resonance observed in \textbf{a}. (\textbf{c})  Experimental spectra of the 1D lattice vibrational heating, indicated here by atom loss due to lower signal:noise in $\sigma$ than \textbf{a}.} 
\label{fig:3_3D_resonance}
\end{figure*}

A spectrum of the cloud  spatial standard deviation after modulation, as a function of modulation frequency, illustrates the heating of the atoms in the near-resonant 3D lattice (Fig.~\ref{fig:3_3D_resonance} \textbf{a}). We see a clear primary peak 
at around $100\,$kHz, with a partially-resolved secondary feature at around $50\,$kHz. Because the 3D lattice is relatively shallow, broad resonances are expected as the atoms explore the full sinusoidal potential, not just the harmonic lattice minima.

We have also determined the expected theoretical trap frequencies in the 3D lattice, allowing for the spread due to lattice intensity variation across the atomic cloud from the Gaussian nature of the input and diffracted beams. For the conditions in Fig.~\ref{fig:3_3D_resonance} \textbf{a} this yields trap frequencies in the range $\nu_{z}=(100\pm 10)\,$kHz and $\nu_{r}=(46\pm5)\,$kHz, where standard deviations are given, also allowing for the spatially varying lattice population due to the initial MOT cloud distribution used for loading. 
In addition, the atoms after optical molasses are usually populated uniformly across the magnetic hyperfine levels. While subject to the circularly  polarised lattice beam \footnote{Required for balanced diffraction from all three sectors of the grating.} the atoms are subject to the theoretical trap frequencies above corrected by a factor $\sqrt{\{2,3,4,5,6\}/4}\approx\{0.71, 0.87, 1.00, 1.12, 1.22\},$ depending on which of the five different $F=2$ $m_F$ states they are in.

The experimental 3D lattice heating frequencies therefore match surprisingly well with 
the expected trap frequencies \footnote{Similar discrepancies between theory and experiment have been reported previously, with experimental frequencies (60-90)\% below the maximum theoretical expectations in several separate experiments Ref.~\cite{parametric_oscillation_Wu2006}.}. We also found that changing the lattice depth affects the resolution of the primary and secondary features in Fig.~\ref{fig:3_3D_resonance} \textbf{a}, with data for a range of five powers in the SM. 
This may be due to atom loss from lower depth lattice sites at lower laser power leading to a remaining trapped atom distribution predominantly occupying the central portion of the lattice structure with a lower spread of trap frequencies.

To further test that the atoms experience a lattice arising due to the dipole force, we measured the primary lattice resonance frequency for a  variety of beam powers $P$, and red detunings $\Delta$. 
By plotting the primary observed resonance frequency as a function of $P/\Delta$, resonance data for different 3D lattice trap depths can be summarised in a single plot, as shown in Fig.~\ref{fig:3_3D_resonance}\textbf{b}. 
 As expected, these data follow a $\nu=A\sqrt{P/\Delta}$ relation, where here the fitted value is $A=34.2\,$kHz$\,($GHz/mW$)^{1/2}$, in good agreement with theory.
 
The measured parametric resonance frequencies in the 1D lattice (Fig.~\ref{fig:3_3D_resonance} \textbf{c}) are $\nu_{z}=200\,$kHz and $\nu_{r}=0.5\,$kHz, which, like in the 3D case, match well with sub-harmonics of the expected trap frequencies, $\nu_{z}=290\,$kHz and $\nu_{r}=0.9\,$kHz. There should be much less spatial variation in trap frequency due to lattice structure in the 1D lattice compared to the 3D lattice, and the atoms are expected to be in the deeper harmonic parts of the lattice, with no trap frequency variation with $m_F$ state. In the 1D case, however, the signal-to-noise of the parametric heating is low, and the (relatively broad) resonances can only be observed via atom number loss. This is due to lower spatial overlap of the 1D lattice with the MOT and hence lower initial atom number  (Fig.~\ref{fig:lifetime}), as well as rapid radial expansion during fluorescence imaging (Fig.~SM2 and discussion in the SM). The 1D lattice beam focal position on the grating surface was optimised for atom number, and given the lattice beam Rayleigh range of $19\,$mm, the beam waist was constant over the entire lattice.  

In this paper we have demonstrated optical lattices in one and three dimensions. However, we note that by loading the MOT atoms into a magnetic and/or optical trap, they can be transported beyond the MOT capture volume, and thus future single-input-beam two-dimensional lattices are within easy reach (Fig.~\ref{fig:fig1_place_holder} \textbf{b}, \textbf{c}).

We have demonstrated experimentally that the GMOT architecture can be extended from a single-input-beam MOT, to both three- and one-dimensional single-input-beam optical lattices on the same chip that is also likely to enable 2D lattices. Because the grating surface, from which all other lattice beams originate, is a node of the electric field the associated beam phases and thereby grating lattices are expected to be much more stable than standard optical lattices derived from multiple centimetre-distant mirrors.

The 3D and 2D lattices arising from our binary grating chip with three-fold symmetry (`tri' chips) naturally generate lattices of a topical \cite{Struck2011,Xu2023,Mongkolkiattichai2023} triangular nature. However, equivalent 3D body-centred-cubic lattices (comprising offset 2D square lattice layers) can be generated using binary gratings based on either four sectors of linear grating (`quad' chips \cite{GMOT1_Nshii2013,Imhof2017,Burrow2021}), or the larger capture volume 2D checkerboard grating option \cite{GMOT1_Nshii2013}. By moving beyond binary gratings into multi-level gratings, it is expected that more exotic lattice geometries may  also be explored \cite{Viebahn2019,Wintersperger2020,Leung2020}. Given that typical optical lattices have beam waists around $100\,\upmu$m, it seems natural to have a GMOT grating region bordered by a wide variety of small grating patterns that the MOT can be transported to, to easily explore different lattice geometries within the same experimental setup. 

We therefore expect our technology is ideally suited for application in a wide range of stable, high-accuracy, portable quantum devices.
 
\begin{acknowledgments}
The authors would like to thank Arthur La Rooij and Oliver Burrow for valuable discussions. 
A.B.\ was supported by a Ph.D. studentship from the Defence Science and Technology Laboratory (Dstl). J.P.M. gratefully acknowledges support from a Royal Academy of Engineering Research Fellowship. We thank the UK EPSRC for funding via grant EP/T001046/1. For the purpose of open access, the
authors have applied a Creative Commons Attribution (CC BY)
licence to any Author Accepted Manuscript (AAM) version
arising from this submission. The dataset is available here \footnote{Dataset DOI to be added.}, and supplementary material which includes Refs.~\cite{Vangeleyn2010,Henderson2020,Arnold2000} can be found here \footnote{Supplementary Material link to be added}.
\end{acknowledgments}

\bibliography{Lattice.bib}

\end{document}